\documentclass[]{zakoproc}
\usepackage{graphicx}
\usepackage{enumitem}
\setlist{nolistsep}

\newcommand{\simgt}%
          {\,\hbox{\lower0.45ex\hbox{$\sim$}%
             \llap{\raise0.45ex\hbox{$>$}}}\,}

\begin{document}

\title{The role of star formation for the galactic dynamo}

\author{Detlef Elstner$^1$  \&  Oliver Gressel$^2$}
\institute{$^1$Leibniz Institut for Astrophysics Potsdam, 
              An der Sternwarte 16, 14482 Potsdam, Germany\\
          $^2$Astronomy Unit, Queen Mary University of London, 
              Mile End Road, London E1 4NS, UK}
\markboth{D. Elstner}{The role of starformation \ldots}

\maketitle

\begin{abstract}
Magnetic field amplification by a fast dynamo is seen in local box
simulations of SN-driven ISM turbulence, where the self-consistent
emergence of large-scale fields agrees very well with its mean-field
description. We accordingly derive scaling laws of the turbulent
transport coefficients in dependence of the SN rate, density and
rotation. These provide the input for global simulations of regular
magnetic fields in galaxies within a mean-field MHD framework. Using a
Kennicutt-Schmidt relation between the star formation (SF) rate and
midplane density, we can reduce the number of free parameters in our
global models. We consequently present dynamo models for different
rotation curves and radial density distributions.
\end{abstract}

\section{Introduction}

Simulations of the ISM in a shearing box domain have shown that
turbulence driven by SNe leads to an amplification of the mean
magnetic field. Using the test-field method (Schrinner
et~al.~\cite{schrinner}), we derived transport coefficients relating
the mean electromotive force to the mean magnetic field
(Gressel~\cite{gressel01}). With these we were able to reproduce the
time behaviour seen in the simulations. Under conditions found in our
own galaxy, and assuming a constant circular velocity, a rotation rate
$\simgt 25\,{\rm Gyr}^{-1}$ was required for the dynamo to work. In
order to further define the turbulence properties as a function of the
star formation rate, rotation and gas density, we analysed a
comprehensive set of direct simulations. Taking these as an input, we
here compute global mean-field maps for a set of different model
galaxies.

\section{The mean-field model} 

Measuring test-field coefficients for a wide set of direct simulations
(Gressel et~al.~\cite{gressel02}, \cite{gressel03}) led to the
following scaling relations for the relevant diagonal term in the
$\alpha$~tensor,
$$\alpha_{\phi\phi}= 2\,{\rm km\,s^{-1}} 
             \left({\sigma \over \sigma_0}\right)^{0.4} 
             \left({\Omega \over \Omega_0}\right)^{0.5}
             \left({\rho \over \rho_0}\right)^{-0.1}\,,      $$
for the (downward) turbulent pumping described by the antisymmetric part
of the $\alpha$~tensor
$$\alpha_{r\phi}= -\alpha_{\phi r} = 8\,{\rm km\,s^{-1}} 
                 \left({\sigma \over \sigma_0}\right)^{0.45}
                 \left({\Omega \over \Omega_0}\right)^{-0.2}
                 \left({\rho \over \rho_0}\right)^{0.3}\,,   $$ 
for the turbulent diffusivity                  
$$\eta= 2\,{\rm kpc^2 Gyr^{-1}}
           \left({\sigma \over \sigma_0}\right)^{0.4} 
           \left({\Omega \over \Omega_0}\right)^{-0.55}
           \left({\rho \over \rho_0}\right)^{0.4}\,,         $$ 
and for the mean vertical outflow velocity 
$$ \bar{u}_z = 15\,{\rm km\,s^{-1}} \left({\sigma \over \sigma_0}\right)^{0.4} 
           {z \over 1{\rm kpc}}\ .                           $$ 
The relations were derived for SF rates, $\sigma$, varying from one
tenth up to the galactic value $\sigma_0=30\,{\rm Myr^{-1} kpc^{-2}}$,
angular velocities between $\Omega_0=25\,{\rm Gyr^{-1}}$ and $10
\Omega_0$ and midplane densities from $0.1\rho_0$ up to
$\rho_0=1\,{\rm cm^{-3}}$. From the simulations, we moreover found a
vertical gradient of the turbulent velocity $\nabla_{\!z}\,u' =
40\,{\rm km\,s^{-1}kpc^{-1}}$ independent of the star formation rate,
density and angular velocity.

\begin{figure}[ht]
 \includegraphics[width=1.\textwidth]{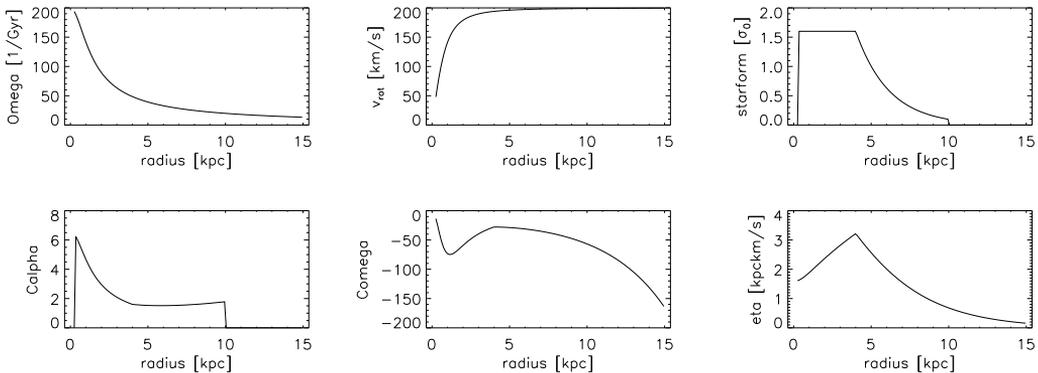}
 \caption{Radial profiles of the rotation curve, the star formation
   rate and the corresponding distribution of the dynamo numbers and
   the turbulent diffusivity for model B1 of Table~1.}
 \label{fig:models}
\end{figure}

We approximate the vertical profiles for the $\alpha$~tensor by a
${\rm sin(2\pi} z/{\rm h})$ curve with a scale height of ${\rm
h=1\,kpc}$. The value of $\eta(z)$ is chosen to be constant for
$-1.5{\rm h}< z < 1.5{\rm h}$ and linearly growing with a slope of one
third outside this range. For simplicity, we assume a constant scale
height within our models. We also neglect the anisotropic part of the
turbulent diffusivity, which seems to be of minor importance for the
current models. The rotation curve is modelled with a Brandt law
$$\Omega=\Omega_0 (1 + (r/{\rm r_\Omega})^2)^{-0.5}.$$ 
Further we modify the vertical wind above ${\rm h}$ by adding a radial
outward velocity of the same size as $\bar{u}_z$.  The wind velocities 
reach values  of 100-200 km/s at z=4kpc, which is an order of magnitude higher 
than in the models of Moss et al.~\cite{moss}.  With these input
parameters, we solve the induction equation
$${\partial \vec{B} \over \partial t}
       = {\rm curl} \left( \vec{u} \times \vec{B} + \alpha \vec{B} 
       - \eta\,{\rm curl} \vec{B}\right)$$ 
in a cylindrical domain with $r < 15 {\rm kpc}$, and of vertical
extent $-4 {\rm kpc} < z < 4{\rm kpc}$. Defining $C_\alpha=\alpha {\rm
h} \eta^{-1}$ and $C_\Omega=\Omega_0 {\rm h^2} \eta^{-1}$, we obtain a
dynamo number $D \equiv C_\alpha C_\Omega \propto \Omega^{2.5}
\rho^{-0.9} \sigma^{-0.4}$. The pitch angle, $p$, can be estimated by
${\rm tan}(p) \propto (C_\alpha/C_\Omega)^{0.5}$, scaling as
$\Omega^{-0.25} \rho^{-0.05} \sigma^{0.2}$. These estimates show that
stronger SF reduces the dynamo number and increases the pitch
angle. It is known that the stationary quadrupole solution of the
$\alpha\Omega$~dynamo exists only in a finite range of the dynamo
number. Because the final strength of the regular field also depends
on the saturation process, this estimate does, however, not provide a
prediction for the final field strength in dependence of the star
formation. Nevertheless, this behaviour still opens the possibility
for radially extended regular magnetic fields. This is because, in an
exponential disc, SF decays much faster with radius than the angular
velocity, and hence the dynamo number may be nearly constant over a
large radial range.
\begin{figure}[ht]
 \includegraphics[width=.5\textwidth]{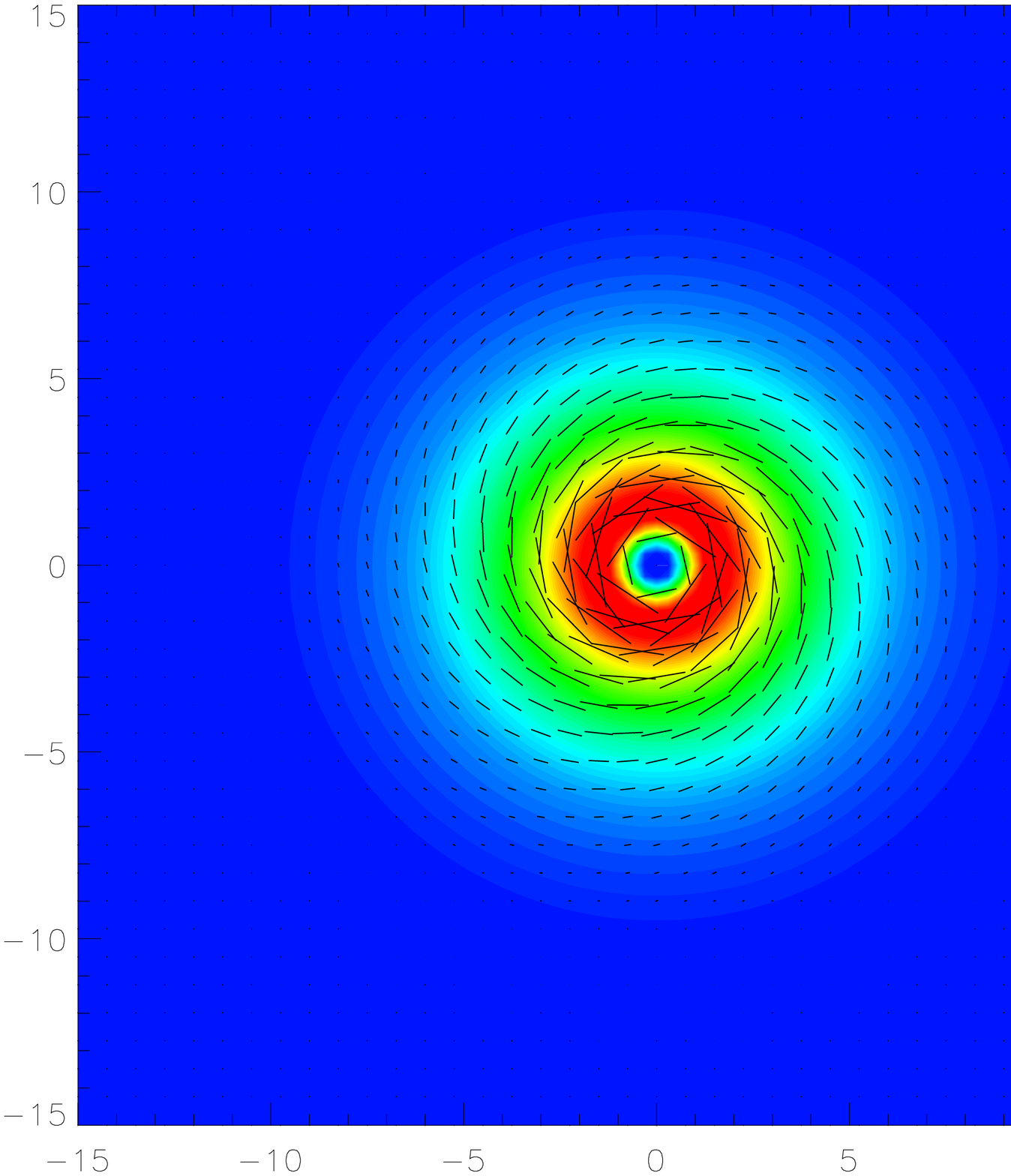}\hfill
 \includegraphics[width=.5\textwidth]{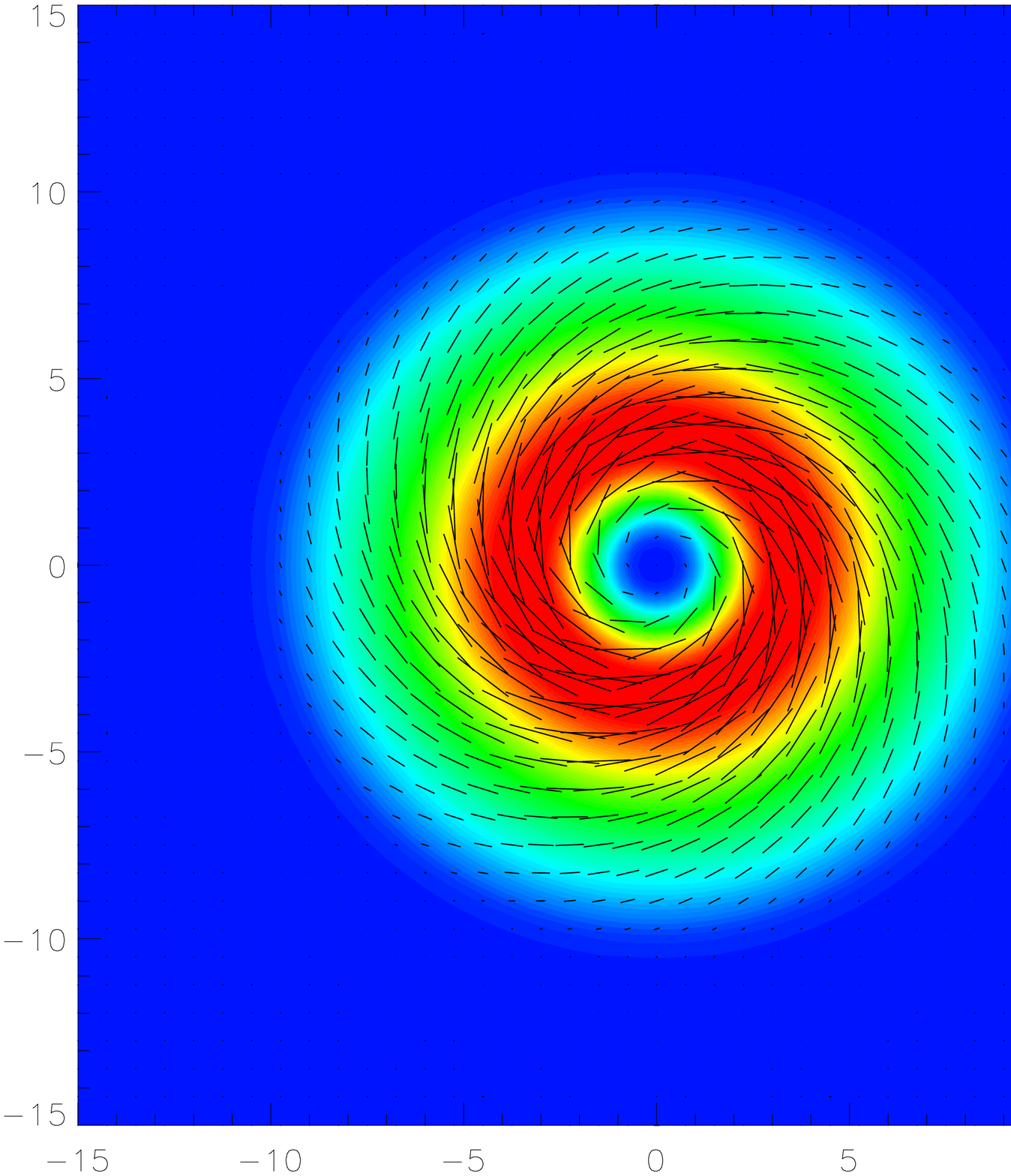}\\
 \includegraphics[width=0.5\textwidth]{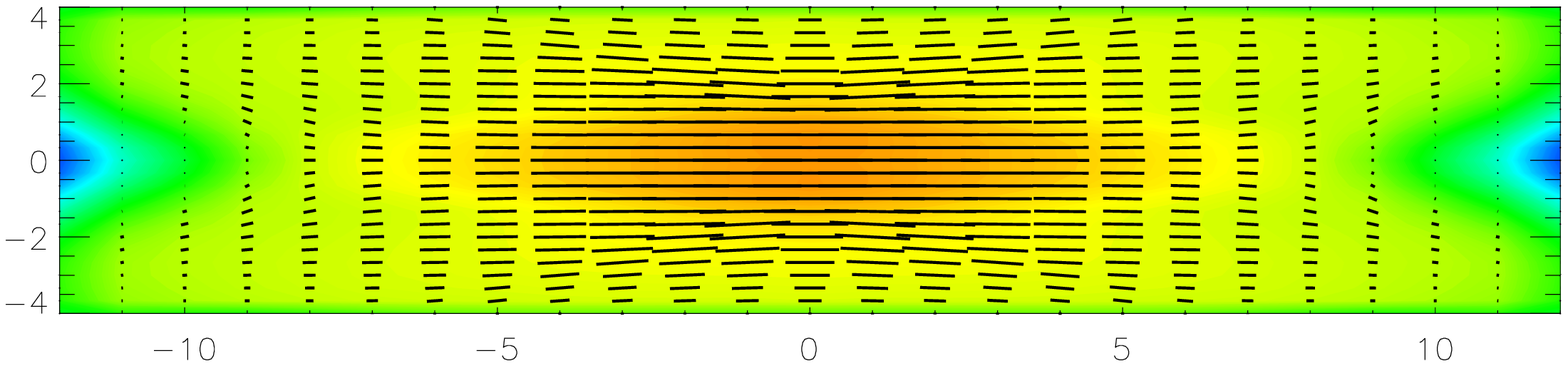}\hfill
 \includegraphics[width=0.5\textwidth]{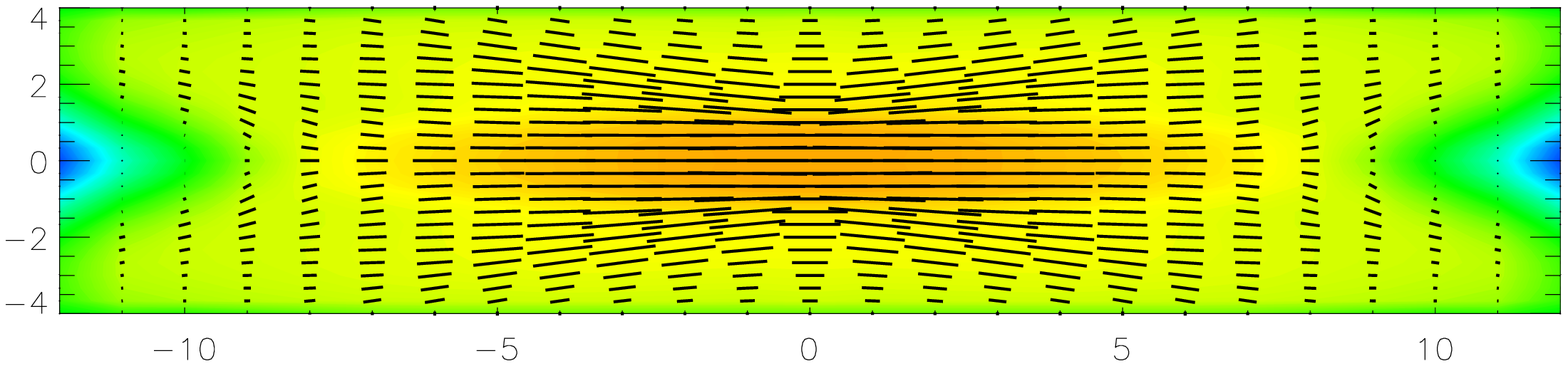} 
 \caption{Total intensity (colour coded) and magnetic field vectors
   for model A2 and A4 of Table~1.}
 \label{fig:polmap}
\end{figure}
Applying a Kenicutt~Schmidt-type law, $\sigma \propto \rho^{1.4}$, we
can specify our galaxy model by a radial density profile, which we
leave constant up to $r_\rho$ and then exponentially decay with a
scale length of $3\,{\rm kpc}$, as seen in Fig.~\ref{fig:models}. For
the nonlinear back-reaction, we use a classical, however anisotropic,
$\alpha$~quenching. While the usual quenching of only the diagonal
terms in the $\alpha$~tensor would lead to solutions with a small
pitch angle, independently quenching the pumping term can also
saturate the dynamo by the increasing field advection from the
wind. In this case, the pitch angle of the kinematic mode can be
preserved (Elstner et~al.~\cite{elstner}).

\section{Results}

The models rely on a crude approximation of the vertical profiles for the turbulent transport coefficients, which still leave some uncertainty in the absolute numbers given in Table~1. Nevertheless, the general trend  agrees well with the predictions form the local dynamo number analysis.   The pitch angle, measured from the magnetic vectors (see Fig.~\ref{fig:polmap}) of the polarisation
map at $r_\rho$, increases slightly with the star formation rate as
predicted by the ratio $C_\alpha/C_\Omega$ above. The growth
times of the order of $100\,{\rm Myr}$ tend to increase with the star formation rate, but there are exceptions (cf. A1 and A2 of Table~1). No large-scale magnetic field amplification was
observed for $\sigma > 10 \sigma_0$ in model B3, and in the weakly
differentialy rotating model A5.  Yet, strong starbursts are usually
not long-lasting events and therefore the dynamo may still operate on
longer time scales. The final field strength is not strongly dependent
on the SF rate, and only the toroidal field is influenced by the
difference in turbulent diffusivity. 
\begin{table}[ht]
\caption{Model parameters and results}
\begin{center}
 \begin{tabular}[]{lccccccccc}\hline
   & $r_\Omega$ & $r_\rho$ & $\sigma$ & 
  D($r_\Omega$) &  $B_\varphi$ & $B_r$ & pitch& atan $\sqrt{C_\alpha/C_\Omega}$  & $t_G$  \\
  & [kpc] & [kpc] & [$\sigma_0$]  &  & 
  [$B_{\rm eq}$]  & [$B_{\rm eq}$] & [deg] & [deg]    & [Gyr]\\[5pt]
  \hline
 A1  & 1   & 7 & 0.4 & 1540 & 1.3  & 0.25 & 12 & 12 & 0.18 \\   
 A2    & 1   & 7 & 1   &  550 & 1.1  & 0.25 & 14 & 14 & 0.16 \\
 A3   & 1   & 7 & 2.6 &  192 & 0.95 & 0.25 & 17 & 17 & 0.25 \\
 A4  & 2.5 & 7 & 1   &   86 & 0.65 & 0.28 & 19 & 19 & 0.35 \\
 A5   & 5   & 7 & 1   &   23 & 0.0001 & 0.0001 & -& 38 & $\infty$\\[5pt]
 B1  & 1   & 4 & 1.6 &  344 & 1.1 & 0.26  &  11& 14    & 0.18    \\
 B2  & 1   & 4 & 4.2 &  121 & 0.9 & 0.23  & 13 & 16     & 0.49  \\
 B3  & 1   & 4 & 11.2 &  43 & 0.0001 & 0.0001 & - & 27 & $\infty$ \\
 B4  & 2.5 & 4 & 1.6  & 54  & 0.6 & 0.25  &  17 & 20 & 0.47\\
  \hline
 \end{tabular}
\end{center}
\end{table}
The inverse dependence of the dynamo action on the SF activity is
mainly due to an enhanced turbulent diffusion. This does not
necessarily increase the turbulent velocity but may equally change the
correlation time. In fact, the preference of magnetic arms
{\it between} the optical arms may be based on this very property of
ISM turbulence (cf. Rohde et al.~\cite{rohde}).

%\section{Conclusions}

The main conclusions drawn from the presented set of simulations are
as follows:\vspace{1ex}
\begin{itemize}
\item[$\bullet$]{SF rate and rotation determine the dynamo -- calling
 for adequate galaxy evolution models.}
\item[$\bullet$]{Low star formation rates favour the dynamo,
 explaining coherent inter-arm fields.}
\item[$\bullet$]{Strong SF may suppress large-scale dynamo action (no
 vertical fields in the centre).}
\item[$\bullet$]{Explaining the radio-FIR relation will require a
 different type of amplification mechanism -- at least for the
 small-scale field.}
\end{itemize}

\acknowledgements{This work was supported by a DFG grant within the
 research unit 1254.}

\end{document}